\documentclass[3p,times]{elsarticle}

\usepackage{ecrc,color}

\volume{00}

\firstpage{1}

\journalname{Acta Astronautica}

\runauth{}

\jid{procs}

\jnltitlelogo{Acta \\Astronautica}

\usepackage{amssymb}

\usepackage[figuresright]{rotating}
\usepackage[cmex10]{amsmath}
\usepackage{enumitem}
\usepackage{array}
\usepackage{fixltx2e}
\usepackage{tabularx}
\usepackage{url}
\begin{document}

\begin{frontmatter}



\dochead{}

\title{Genetic Algorithm Based Robust and Optimal Path Planning for Sample-Return Mission from an Asteroid on an Earth Fly-By Trajectory}


\author{Sean Fritz \thanks{sean.fritz@sjsu.edu} and 
Kamran Turkoglu\thanks{kamran.turkoglu@sjsu.edu} }

\footnote{Graduate student, Control Science and Dynamical Systems (CSDy) Laboratory, Aerospace Engineering, San Jose State University, sean.fritz@sjsu.edu}
\footnote{Assistant Professor, Aerospace Engineering, San Jose State University, kamran.turkoglu@sjsu.edu}

\address{San Jose State University, Aerospace Engineering, San Jose, 95192, CA USA}
 
\begin{abstract}
This paper investigates a systematic mission design of robust and optimal orbital transfer maneuvers for a sample return mission from an asteroid. In this study, an interplanetary space flight mission design is established to obtain the minimum \(\Delta V\) required for a rendezvous and sample return mission from an asteroid. Given the initial (observed) conditions of an asteroid, a (robust) genetic algorithm is implemented to determine the optimal choice of \(\Delta V\) required for the rendezvous. Robustness of the optimum solution is demonstrated through incorporated bounded-uncertainties in the outbound \(\Delta V\) maneuver via genetic fitness function. The improved algorithm results in a solution with improved robustness and reduced sensitivity to propulsive errors in the outbound maneuver. This is achieved over a solution optimized solely on \(\Delta V\), while keeping the increase in \(\Delta V\) to a minimum, as desired. Outcomes of the analysis provide significant results in terms of improved robustness in asteroid rendezvous missions. 
\end{abstract}


\end{frontmatter}
 

\section{Introduction}
Asteroid rendezvous with man-made objects are a very recent human endeavor. One of the main reasons of such a pursuit is that asteroids and comets give us a window into the past and hold very valuable information about the formation of the universe, and how the solar system was created \cite{Glassmeier}.  If an asteroid were to fly-by earth, we would have a unique window of opportunity to conduct a scientific mission to deduce creation of the solar system. The first spacecraft to orbit and land on an asteroid was the Near Earth Asteroid Rendezvous (NEAR) Shoemaker spacecraft, which orbited 433 Eros in February 2000 and landed on 433 Eros in February 2001 \cite{Holdridge}. One of the first successful sample return missions was ISAS MUSES-C. In this mission, an asteroid sample from Nereus (4460) was returned to Earth. The MUSES-C spacecraft successfully used the ION thruster propulsion system for the inter-planetary cruise phase of its mission \cite{Kawaguchi}.

The trajectory design for interplanetary rendezvous missions to asteroids is quite 	complicated and often derived by first dividing the mission at hand into segments, such as an Earth-Asteroid leg, an Asteroid-Asteroid leg and Asteroid-Earth leg, and then reconnecting these segments after they are evaluated against some mission constraints, such as payload mass \cite{Morimoto}. Morimoto M. et al provided a preliminary study using a genetic algorithm approach to solving the problem of trajectory design and it has the potential to eliminate the enormous calculation time required by brute force optimization \cite{Morimoto}. \cite{bulirsch1992optimal} approached to the planning of an asteroid rendezvous mission as a multipoint boundary value problem which is solved by the multiple shooting method. In their study, \cite{wall2009shape} tackled mission planning framework of their (generally speaking) trajectory optimization problem with a hybrid approach, which combines "a shape-based`` non-linear programming problem with genetic algorithms and extended it further in \cite{wall2009genetic}.  \cite{park2005optimal} presents a methodology for in-direct optimization methods of the mitigation trajectories in three-dimensional space by using a spacecraft with variable thrust. In his work, \cite{izzo2007optimization} provides very useful insights regarding the optimization of interplanetary trajectories for impulsive and continuous asteroid deflection. In another intersting recent effort, \cite{pontani2012particle} initiated swarm optimization algorithms to tackle trajcetory planning problem. \cite{colombo2009optimal_differential} tackled the problem from the differential dynamic programming perspective.

In all of those valuable studies, several methods have been discussed to tackle the problem from different perspectives, but one major drawback in those studies were the highly uncertain space environment and its effects on the mission planning. Especially, when the distances that are traveled by those probes are taken into account (i.e. anywhere from 1e6 [km] to light years) the significance of robustness analysis and robust optimal solutions becomes more important. Therefore, in this study, different than the literature, we aim to look into the robustness and robust optimization methodologies of the orbital mission and trajectory planning aspect of a problem to collect a sample from a near-Earth asteroid and return it back to Earth. The goal of this study is to lay down a robust, systematic procedure of sample-collection from an asteroid, from launch to landing with the optimal trajectory and maneuvers (i.e. thruster firings). This optimal set-up is achieved through the powerful nature of genetic algorithms and is implemented through a domain of uncertainties, to demonstrate the robust optimal solutions and the feasible sets. This paper differs from past missions in a way that the mission objectives will be laid down for every single step, and robustness analysis is conducted for the first leg of the mission. The goal is for the spacecraft to rendezvous with \emph{any given} hypothetical asteroid on a hyperbolic fly-by trajectory around Earth\cite{Curtis}, land and deploy a probe to mine (said) asteroid for a given amount of time, launch from the asteroid`s surface, and return to Earth. This paper specifically concentrates on the trajectory optimization, robustness analysis, successful orbital transfers of the spacecraft to \& from the asteroid and feasible domain analysis of such mission. 


The robustness analysis was improved over the previous method by incorporating the results of small perturbations in the outbound burn of the optimum solution back into a second application of the genetic algorithm for optimization. The optimization program attempts to strike a balance between the \(\Delta V\) of the mission and minimizing the impacts of propulsive errors. This results in a solution that is more robust against small perturbations in the outbound propulsive burn. This robustness does come at a small cost to \(\Delta V\) as a faster transfer results in less time for the small propulsive errors to compound throughout the trajectory.

The outline of the paper is as following: Section-II introduces the genetic algorithm approach on a sample mission, whereas in Sections III-V the complete systematic procedure is laid out. With the final discussions and remarks in the final section, the paper is concluded.

\section{Orbital Trajectory Optimization via Genetic Algorithms}
In this study of a sample collection mission to a hypothetical asteroid on a fly-by trajectory of Earth some assumptions are made to bound the problem, without the loss of generality. It is assumed that the minimum time spent in the vicinity of the asteroid is bounded, and enforced as a constraint. This helps to define a realistic time period for conducting planned operations.  It is also assumed that the rendezvous can be accomplished at any point in the asteroid's orbit. This permits for a more open-ended timing search for initial transfer orbit, and (possibly) allowing for a smaller \(\Delta\)V requirement. The asteroid's mass is going to be neglected in two-body motion analysis, which simplifies the problem without introducing a large amount of error due to the relatively small mass of the asteroid. Optimization can be performed in two rounds: first to find the optimum trajectory where the mission objective is to minimize the \(\Delta V\) through optimal trajectory and maneuvers, and a second round to adjust this trajectory to improve targeting of the asteroid on the outbound leg of the mission. This will be achieved using a genetic algorithm due to their effective nature when exploring large, non-linear search domains.

\subsection{Genetic Algorithm Approach for Minimum Fuel Burn Trajectory Design}
Genetic algorithms are a stochastic optimization method and for a given proper problem set-up and parameter bounding, they are known to provide near-global optimum solutions. \nocite{Eiben} Within this context, in the following sections we provide a glimpse to the mission planning through genetic algorithm formulation.
\subsubsection{Mission Concept and Analysis}
For the optimal trajectory solution of the problem, the problem space is divided into various different phases:
\begin{itemize}
\item Boost to a parking Low Earth Orbit by the launch vehicle
\item Conduct a plane change in the parking Low Earth Orbit to make the rendezvous coplanar (if necessary)
\item Rendezvous course to the asteroid
\item Landing on the asteroid
\item Mission execution (data gathering, sample collection ... etc.)
\item Takeoff from the asteroid
\item Return course to Earth
\item Reentry and Landing on the Earth
\end{itemize}

Although the launch, re-entry and landing on earth segments are also important sectors of such a mission, analysis of those parts will be omitted in this paper and the main emphasis will be on optimal trajectory and path planning calculations. 

In order to obtain an optimal solution (trajectory) that consumes minimal fuel over the phases not involving Launch and Reentry, a genetic algorithm approach is utilized. In order to simplify the setup for genetic algorithm, following set of phases are used without losing the generality of the proposed mission concept:
\begin{enumerate}[label=\arabic*.]
\item Conduct plane change operation in the parking Low Earth Orbit to bring the spacecraft into the same plane of the asteroid
\item Starting from the resulting Low Earth Orbit (LEO1), perform a burn to transfer the spacecraft onto a rendezvous trajectory.
\item Perform a burn to rendezvous with asteroid.
\item Execute assigned mission (data gathering, sample collection ... etc.) on the asteroid.
\item Depart from the asteroid (and/or the park orbit around the asteroid) and perform a burn to put the spacecraft on a return trajectory to Earth.
\item Perform a burn to place the spacecraft into a Low Earth Orbit (LEO2)
\end{enumerate}

\subsection{Implementation}
In this section, for the sake and completeness of the analysis, the general structure of a genetic algorithm and its components is visited.
\begin{description}
\item[Chromosome] A chromosome is composed of parameters called genes that designate aspects of all the above phases in a sequence. For the existing problem set-up, the chromosomes are comprised of the following 8 genes:\\

[\emph{dt0; leo1\_r; leo1\_nu; dt1; dt2l leo2\_r; leo2\_nu; dt3}]\\

The detailed information of each parameter presented here is given in Table \ref{tab:Parameters}.
\item[Initial population size] An initial random population of 10 trajectories is chosen. It has been observed that a population of 100 trajectories also converges to a similar optimal solution. However, due to high run time of the Genetic Algorithm, a population size of 10 is preferred.
\item[Fitness of Chromosome] The fitness of each chromosome is determined by calculating the \(\Delta V\) required in each of the following maneuvers:
\begin{enumerate}[label=\arabic*.]
\item Plane change from launch site latitude to that of the asteroid.
\item Firing to set off on rendezvous trajectory from LEO\_1[\(\Delta V_1\)]
\item Firing to match the asteroid velocity at the rendezvous point [\(\Delta V_2\)]
\item Firing to match the return trajectory velocity at the end of the mission [\(\Delta V_3\)]
\item Firing to reach LEO\_2 from the return trajectory [\(\Delta V_4\)]\\
\end{enumerate}

\begin{table*}
\centering
\caption{Parameters Operated on by the Genetic Algorithm, their Initial Ranges and Mutation Limits.}
\label{tab:Parameters}
\begin{tabularx}{\textwidth}{l X l l}
\hline \hline\\[-1.5ex]
Parameter & Description & Initial Range & Standard Deviation of Mutation\\[0ex] \hline \hline \\[-1.5ex]
dt0 & Time before/after the observation point given in problem statement before taking off on a rendezvous trajectory to the asteroid. & 0 to 35 hours & 3 hours\\[0ex] \hline \\[-1.5ex]
leo1\_ r & Radius of the position in LEO 1 from where the spacecraft takes off for the asteroid. & 6478 - 6878 km & 10 km\\ [0ex] \hline \\[-1.5ex]
leo1\_nu & True Anomaly of the position in LEO 1 from where the spacecraft takes off for the asteroid. & 0 - 2π & 1 rad\\ [0ex] \hline \\[-1.5ex]
dt1 & Time of travel in the rendezvous trajectory to the asteroid. & 1 - 96 hours & 3 hours\\ [0ex] \hline \\[-1.5ex]
dt2 & Time of stay on the asteroid for the sample collection mission. & 6 - 24 hours & 1 hour\\ [0ex] \hline \\[-1.5ex]
leo2\_r & Radius of the position in LEO 2 to which the spacecraft returns. & 6448 km radius & N/A\\ [0ex] \hline \\[-1.5ex]
leo2\_nu & True Anomaly of the position in LEO 2 to which the spacecraft returns. & 0 - 2π & 1 radian\\ [0ex] \hline \\[-1.5ex]
dt3 & Time of travel in the return trajectory to Earth. & 1 - 96 hours & 3 hours\\ [0ex] \hline \hline \\[-1.5ex]
\end{tabularx}
\end{table*}

In order to obtain the four \(\Delta V\)'s listed above, the following steps are undertaken:
\begin{enumerate}[label=\arabic*.]
\item Kepler's universal variable solution \cite{Bate} is used to determine the asteroid's location and velocity at the time of rendezvous (dt0 + dt1).
\item With the asteroid position at rendezvous and the position of the spacecraft in LEO1 known, the universal variable approach to Gauss' problem \cite{Bate} is used to determine the necessary velocities at each position for a connecting trajectory.
\item \(\Delta V_1\) can now be obtained by simple vector manipulation of the LEO circular and LEO departure velocities at the LEO1 position.
\item \(\Delta V_2\) can be easily obtained by vector algebra using the asteroid and rendezvous trajectory velocities at the rendezvous point.
\item For the general application of the problem to different set-ups, the departure point and velocity of the asteroid could be easily determined using Kepler's solution once again at the time of departure (dt0+dt1+dt2). 
\item Gauss' problem solution \cite{Bate} then could be used a second time to determine a trajectory that will connect the asteroid departure point and the LEO2 position, resulting in spacecraft velocities at each point. 
\item \(\Delta V_3\) is calculated through velocities at the departure point.
\item \(\Delta V_4\) is calculated through velocities at the LEO2 position.
\item Total \(\Delta V\) is calculated as the sum of \(\Delta V_1\), \(\Delta V_2\) and \(\Delta V_3\). \(\Delta V_4\) is a construct to penalize trajectories that would be too steep for reentry.
\end{enumerate}

These steps can are summarized in Figures \ref{fig:ga_flow} and \ref{fig:fitness_flow}.
\begin{figure}[!t]
\centering
\includegraphics[width=5in]{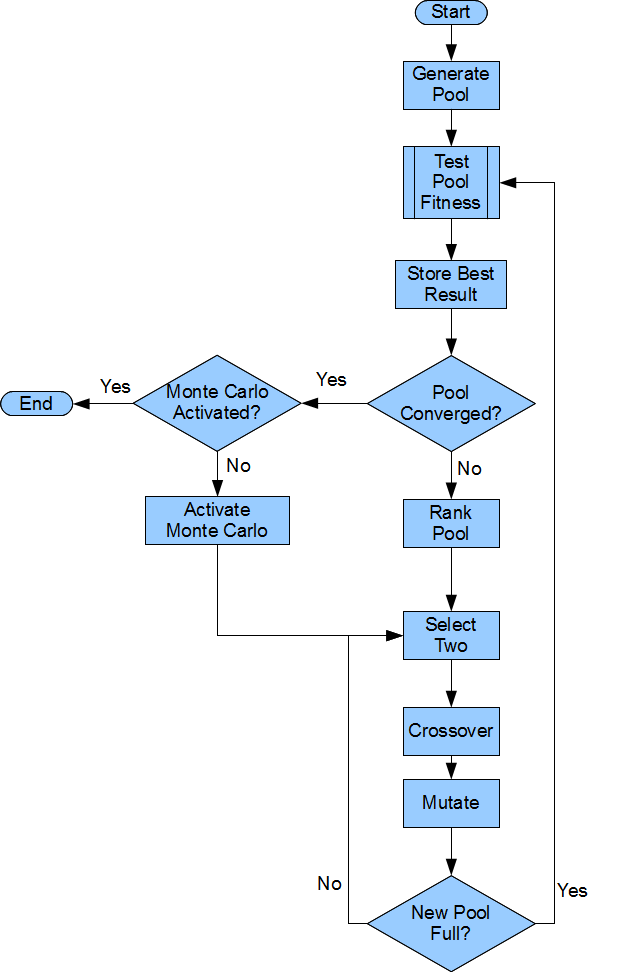}
\caption{Genetic Algorithm Flowchart}
\label{fig:ga_flow}
\end{figure}

\begin{figure}[!t]
\centering
\includegraphics[width=5in]{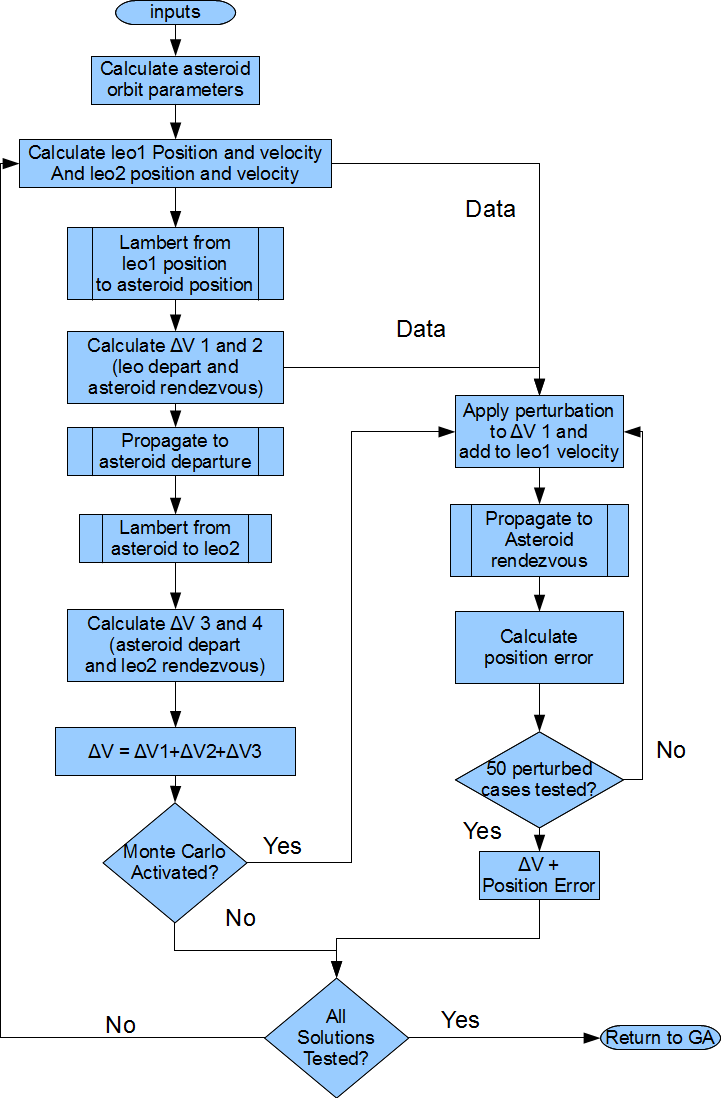}
\caption{Flowchart of the fitness function including Monte Carlo simulations}
\label{fig:fitness_flow}
\end{figure}

In addition to the \(\Delta V\), another aspect of fitness is included in the fitness parameter. In recognition of the fact that uncertainties will be present in the maneuvers, the fitness parameter includes the effects of such uncertainties once the the near-optimal solutions are found. The uncertainties are modeled as Gaussian error distributions with a mean of zero and 3-sigma values of 0.1\% for \(\Delta V\) magnitude and 0.5\(^{\circ}\) for angular error. Then, they are used in a Monte Carlo approach wherein the new outbound trajectory is numerically integrated for the time duration specified by dt1 in each solution. The the final distance from the target from each Monte Carlo set is RSS'd into a single value for the given solution. The final fitness value is determined by adding the total \(\Delta V\) and the RSS'd position error with a higher weight factor being given to the \(\Delta V\) value so as not to let the scatter effect from the burn inaccuracies override the selection process described in the next piece.

\item[Selection criteria for next generation] A roulette wheel scheme with more weight given to better solutions is used to pick the next generation candidates from current generation. A random number is generated in the range of (0, 1), and the first chromosome with a weight, based on its fitness, more than that random number is selected for inclusion in the next generation, after the following modifications are applied.
\item[Crossover] A pair of chromosomes chosen for next generation are mated or crossed over to generate the offspring for the next generation. The process involves stepping through the each gene in the chromosomes and generating a random number between (0, 1). If the number is greater than 0.3, the parent chromosomes swap all genes after the current one with each other to produce two new chromosomes for the next generation.
\item[Mutation] The offspring generated from the crossover algorithm are now modified to introduce new gene values that were not previously generated. For each gene in the chromosomes a random number is generated between (0, 1). If it is less than 0.05, that particular gene has a random value added to it. This random value is generated from a normal distribution whose range is unique to that gene.
\item[Genetic Algorithm Termination Condition] For the purposes of termination of the Genetic Algorithm, a convergence of total fitness, within a specified tolerance, for over 30 generations or surpassing a maximum number of generations is used.
\end{description}

\subsection{Results from the Genetic Algorithm}
The genetic algorithm yielded the following optimized chromosomes, where the results prior to the robust optimization were 
$$ [ -370150.2;~~~ 6878;~~~ 0.5960;~~~ 360000;~~~ 21600;~~~ 6428.5;~~~ 0.1663;~~~ 120782.4 ]$$
and after robust optimization became
$$[ -375842.1;~~~ 6874.6;~~~ 0.6130;~~~ 359719;~~~ 22169.5;~~~ 6428.5;~~~ 5.9526;~~~ 125255.5 ]$$
which correspond to the following solution vector
\noindent
\begin{center}
[\emph{dt0;~~~ leo1\_r;~~~ leo1\_nu;~~~ dt1;~~~ dt2;~~~ leo2\_r;~~~ leo2\_nu;~~~ dt3}]
\end{center}

At this point, it important to note that obtained robust optimization results provide guarantees for optimal solutions within the given uncertainty bounds, which were explained in previous section. From that perspective the overall contribution is significant.

\noindent
Here, once again, the details of which are provided in Table \ref{tab:Parameters}.
Table \ref{tab:Optimized} lists the genes of the winning chromosome.

\begin{table}[!h]
\centering
\caption{Optimized Chromosome Parameters}
\label{tab:Optimized}
\begin{tabularx}{\columnwidth}{X X X}
\hline \hline\\[-1.5ex]
 & \multicolumn{2}{c}{Value}\\ 
\cline{2-3}\\
 Paramater & Before MC & After MC\\ \hline \hline \\[-1.5ex]
dt0 & 35 hours & 34.997 hours\\ \hline \\[-1.5ex]
leo1\_ r & 6685 km & 6702.1 km\\ \hline \\[-1.5ex]
leo1\_nu & 6.0971 radians & 6.0157 radians\\ \hline \\[-1.5ex]
dt1 & 13.6 hours & 7.8 hours\\ \hline \\[-1.5ex]
dt2 & 6 hours & 6.12 hours\\ \hline \\[-1.5ex]
leo2\_r & 6438 km & 6438.4 km\\ \hline \\[-1.5ex]
Leo2\_nu & 3.4576 radians & 3.8780 radians\\ \hline \\[-1.5ex]
dt3 & 100 hours & 99.87 hours\\ \hline \hline \\[-1.5ex]
\end{tabularx}
\end{table}

On the return trajectory the closest approach to Earth (or Perigee point \{leo2\_r = 6428.5 [km], 50.5 [km] altitude\}) is well within the noticeable atmosphere of the Earth and hence the drag will draw the spacecraft into ballistic reentry and landing on Earth. For this reason, the result for \(\Delta V_4\) is ignored for the purposes of sizing the spacecraft, when necessary. It is also assumed that the launch vehicle is capable of placing the spacecraft in a parking orbit LEO\(_1\). It is estimated that ~10 km/s of \(\Delta V\) is required to boost to LEO\(_1\). The \(\Delta V\) for each of these maneuvers and for the total mission is listed in Table \ref{tab:DeltaV}.

\begin{table}[!h]
\centering
\caption{\(\Delta V\) Maneuver List for Asteroid Rendezvous}
\label{tab:DeltaV}
\begin{tabularx}{\columnwidth}{l X l l}
\hline \hline\\[-1.5ex]
 & & \multicolumn{2}{c}{\(\Delta V\) (km/s)}\\
\cline{3-4}\\[-1.5ex]
Phase & Maneuver & Before MC & After MC\\ \hline \hline \\[-1.5ex]
Maneuver 1 & Launch to LEO1 & 10 & 10\\ \hline \\[-1.5ex]
Maneuver 2 & Boost to rendezvous trajectory & 3.23 & 3.06\\ \hline \\[-1.5ex]
Maneuver 3 & Fire to match the velocity of the asteroid & 0.65 & 1.33\\ \hline \\[-1.5ex]
Maneuver 4 & Fire for guided landing on the asteroid & 0.19 & 0.19\\ \hline \\[-1.5ex]
Maneuver 5 & Fire to takeoff from the asteroid & 0.31 & 0.31\\ \hline \\[-1.5ex]
Maneuver 6 & Fire to get into return trajectory & 2.04 & 1.75\\ \hline \hline \\[-1.5ex]
Total &  & 16.42 & 16.64\\ \hline \hline
\end{tabularx}
\end{table}

\subsection{Trajectory and Orbital Parameters}
\begin{figure}[!t]
\centering
\includegraphics[width=\columnwidth]{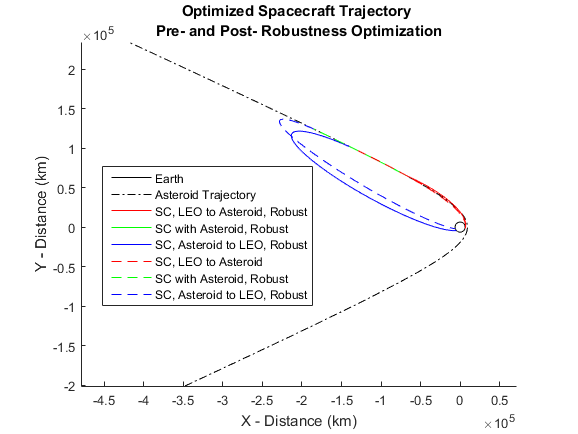}
\caption{Optimized Trajectory in Perifocal Frame}
\label{fig:Optimized}
\end{figure}

At the end of the analysis, obtained optimal path of the entire mission trajectory is shown in Figure \ref{fig:Optimized}.
where the corresponding orbital Parameters are listed in Table \ref{tab:OrbitalPre} and Table \ref{tab:OrbitalPost}, respectively.
\begin{table}[!h]
\centering
\caption{Pre-MC Orbital Parameters}
\label{tab:OrbitalPre}
\begin{tabularx}{\columnwidth}{X X X X}
\hline \hline\\[-1.5ex]
Orbit & Semi-major axis (km) & Eccentricity & Time (hrs)\\ \hline \hline \\[-1.5ex]
Rendezvous trajectory & \(-523199.5\) & 1.0128 & \(\infty\)\\ \hline \\[-1.5ex]
Asteroid trajectory & \(-93416.4286\) & 1.0989 & \(\infty\) \\ \hline \\[-1.5ex]
Return trajectory & 133090.5 & 0.9916 & 134.2 \\ \hline \hline
\end{tabularx}
\end{table}

\begin{table}[!h]
\centering
\caption{Post-MC Orbital Parameters}
\label{tab:OrbitalPost}
\begin{tabularx}{\columnwidth}{X X X X}
\hline \hline\\[-1.5ex]
Orbit & Semi-major axis (km) & Eccentricity & Time (hrs)\\ \hline \hline \\[-1.5ex]
Rendezvous trajectory & 118863.9 & 0.9437 & 113.3 \\ \hline \\[-1.5ex]
Asteroid trajectory & \(-93416.4286\) & 1.0989 & \(\infty\) \\ \hline \\[-1.5ex]
Return trajectory & 122831.2 & 0.9819 & 119.0 \\ \hline \hline
\end{tabularx}
\end{table}
Further investigation was also conducted into the tradeoff between robustness and optimizing the total \(\Delta V\) cost of the mission. Multiple iterations of incorporating a Monte Carlo simulation were iterated before settling on a final methodology for improving robustness of the solution.
\subsubsection{Monte Carlo on Every Trial Solution}
An attempt was made to apply the Monte Carlo routine to every potential solution as it was tested in the fitness function. This allowed the genetic algorithm to operate on a combination of the \(\Delta V\) for the trajectory and the solution sensitivity to perturbations. Instead of operating uniformly across the inputs, the perturbations were implemented as a normally distributed randomized error with a maximum 0.1\% error on the magnitude and a \(0.5^\circ\) offpointing of the first (outbound) \(\Delta V\) maneuver. This first maneuver was deemed the most critical as it is responsible for setting the spacecraft on a rendezvous trajectory to the asteroid. 50 Monte Carlo simulations were performed on each trial solution. This resulted in 50 position errors at the end of the first leg of the mission. These errors were RSS'd into a single value that represented the tendency of that solution to scatter in the presence of propulsive perturbations. This "scatter" is the difference between the calculated position of the asteroid at the time of rendezvous and the numerically integrated final position of the spacecraft when randomized velocity and pointing errors were included in the outbound burn.

The optimization process attempts to center and tighten the scatter on the target (the position of the asteroid at the time of rendezvous). This comes at the expense of increased \(\Delta V\). This optimization step would be hindered greatly if the scatter results are not linearly related to the induced perturbations, hence the small perturbations used.

This was added to the \(\Delta V\) per  Eq.(\ref{eq:fitness}) below. Knowing that the scatter value of a solution would be improved with faster travel times and that a large initial burn would improve travel time at a severe cost to \(\Delta V\), a lower weighting was given to the scatter tendency.

\begin{equation}
\label{eq:fitness}
fitness=0.75*\Delta V+0.25*scatter
\end{equation}

Another issue in this approach is the fact that \(\Delta V\) is measured in km/s and is typically less than 10 while the scatter value is measured in km and is often in the range of \(10^5\) to \(10^6\). To keep the values to the same order of magnitude, Eq.(\ref{eq:fitness}) was modified as shown in Eq.( \ref{eq:fit_mod}).

\begin{equation}
\label{eq:fit_mod}
fitness = 0.75*\Delta V+0.25*\frac{scatter}{10^6}
\end{equation}
Although, the analysis portion is crucial, we would like to note that this approach introduced a computational burden in which the computing time was slowed \%10.

At this phase, we also introduced a method of delaying the Monte Carlo implementation until the solutions began to converge. This method relied on calculating the ''spread'' of the solutions through the use of Eq.(\ref{eq:spread}).

\begin{equation}
\label{eq:spread}
spread=\frac{max. result-min. result}{min. result}
\end{equation}

Once the spread was calculated to be below a certain threshold (in our case, 0.15 (i.e. \%15) or less), it is assumed that the solutions were converging and the Monte Carlo code could be enabled. In practice, this spread value was achieved for the first time after about 10 to 15 generations. With the non-Monte Carlo code requiring about 140 generations at a minimum to converge, this threshold approach did not allow for an appreciable speed increase in calculation.

In the proposed methodology, to get around the computational barrier, the approach was altered once again by folding in the Monte Carlo simulation results into the optimization process after an ideal solution was developed. This is in recognition that the range of perturbations is relatively small and the final result should be near the ideal solution regardless of when the perturbations are taken into account by the optimization process.

\begin{figure}[!t]
\centering
\includegraphics[width=0.9\columnwidth]{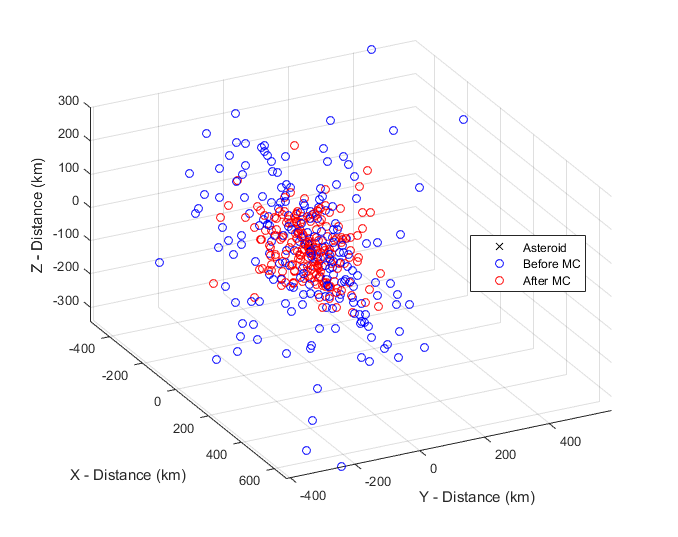}
\caption{Effect of Outbound Burn Perturbations Before and After Robustness Improvement}
\label{fig:Scatter}
\end{figure}

The results of incorporating the Monte Carlo perturbations into the optimization process can be seen in the difference of the \(\Delta V\) results of the ideal trajectory and the ''improved robustness'' trajectory as listed in Table \ref{tab:DeltaV}. It can also be seen in the reduced statistical scatter upon reaching the vicinity of the asteroid as shown in Figure \ref{fig:Scatter}, which plots one of the novel contributions of the paper. The points corresponding to the improved solution are grouped tighter around the center indicating less susceptibility to the applied perturbations, and are depicted with \textcolor{red}{red}. The average distance off-target prior to Monte Carlo improvement was about 235 km off from the center of the asteroid, while the average distance after Monte Carlo feedback was about 110 km, after the robust and optimal path planning. The cost of this improved targeting  manifests as an increase of \(\Delta V\) of about 1 km/s requiring the spacecraft to carry more maneuvering fuel.  The trade-off metric for the associated cost in improvement is worthwhile to investigate and would depend on the needs of the mission itself and the amount of reserves (or margin allocated) to fuel usage during the design of the spacecraft.

This method in improving the \emph{targeting} shows the versatility and robustness of the genetic algorithm in its adaptability to changing priorities. This suggests that a genetic algorithm could be walked through a problem in which a basic solution is found and then improved upon in steps as secondary factors are included in the design of the trajectory.

\section{Conclusions}
This paper has presented a method to determine robust and optimal \(\Delta V\) requirements to rendezvous an asteroid passing Earth on a hyperbolic trajectory. The genetic algorithm based robust optimization program attempts to strike a balance between the $\Delta V$ of the mission and minimizing the impacts of propulsive errors. This results in a solution that is more robust against perturbations (and uncertainties) in the outbound propulsive burn. This approach does come with a small trade-off on $\Delta V$ as faster transfers resulting in less time for the small propulsive
errors to compound throughout the trajectory. The average
distance off-target prior to Monte Carlo has improved $\sim \%50$ after the robust and optimal path planning. These results show the clear benefit and effectivesness of the proposed robustness analysis in improving rendezvous missions with asteroids.


\section*{Acknowledgements}
Authors would like to thank Aaron Mazzulla, Alexander Carlozzi, Dhathri Somavarapu and Zachary Pirkl from San Jose State University for their prior contributions to the study.

\bibliographystyle{elsarticle-num}
\bibliography{Bibliography}







\end{document}